\documentclass{vgtc}            
\newcommand*{\doi}[1]{\href{http://dx.doi.org/#1}{doi: #1}}

\ifpdf%                         
\pdfoutput=1\relax              
\usepackage{graphicx}           
\DeclareGraphicsExtensions{.pdf,.png,.jpg,.jpeg} 
\else%                                 
\usepackage{graphicx}                
\DeclareGraphicsExtensions{.eps}     
\fi%

\graphicspath{{figures/}{pictures/}{images/}{./}} 

\usepackage{arydshln}

\usepackage{microtype}                
\PassOptionsToPackage{warn}{textcomp}  
\usepackage{textcomp}                  
\usepackage{mathptmx}                  
\usepackage{times}                     
         
\usepackage{cite}                      
\usepackage{tabu}                      
\usepackage{booktabs}

\usepackage{array}
\usepackage{pifont}

\onlineid{0}

\vgtccategory{Research}

\vgtcpapertype{Position Paper}

\usepackage{color}
\usepackage[usenames,dvipsnames,table]{xcolor}

\vgtcinsertpkg

\title{Analyzing Visual Mappings\\ of Traditional and Alternative Music Notation}

\DeclareMathSymbol{\ast}{\mathbin}{symbols}{"03}

\author{Matthias Miller\thanks{e-mail:lastname@dbvis.inf.uni-konstanz.de} %
\and Johannes H\"au{\ss}ler$^*$ %
\and Matthias Kraus$^*$
\and Daniel Keim$^*$
\and Mennatallah El-Assady$^*$}
\affiliation{\scriptsize University of Konstanz}
     
\shortauthortitle{? \MakeLowercase{\textit{et al.}}: MusicVIZ: Analyzing Visual Mappings of Traditional and Alternative Music Notation}

\abstract{

	In this paper, we postulate that combining the domains of information visualization and music studies paves the ground for a more structured analysis of the design space of music notation, enabling the creation of alternative music notations that are tailored to different users and their tasks. 
	Hence, we discuss the instantiation of a \textit{design and visualization pipeline for music notation} that follows a structured approach, based on the fundamental concepts of information and data visualization.
	This enables practitioners and researchers of digital humanities and information visualization, alike, to conceptualize, create, and analyze novel music notation methods. 
	Based on the analysis of relevant stakeholders and their usage of music notation as a mean of communication, we identify a set of relevant features typically encoded in different annotations and encodings, as used by interpreters, performers, and readers of music. 
	We analyze the visual mappings of musical dimensions for varying notation methods to highlight gaps and frequent usages of encodings, visual channels, and Gestalt laws. 
	This detailed analysis leads us to the conclusion that such an under-researched area in information visualization holds the potential for fundamental research. 
	This paper discusses possible research opportunities, open challenges, and arguments that can be pursued in the process of analyzing, improving, or rethinking existing music notation systems and techniques.

} % end of abstract


\teaser{
	\vspace*{-0.5cm}
	\centering
	
	\includegraphics[width=\linewidth]{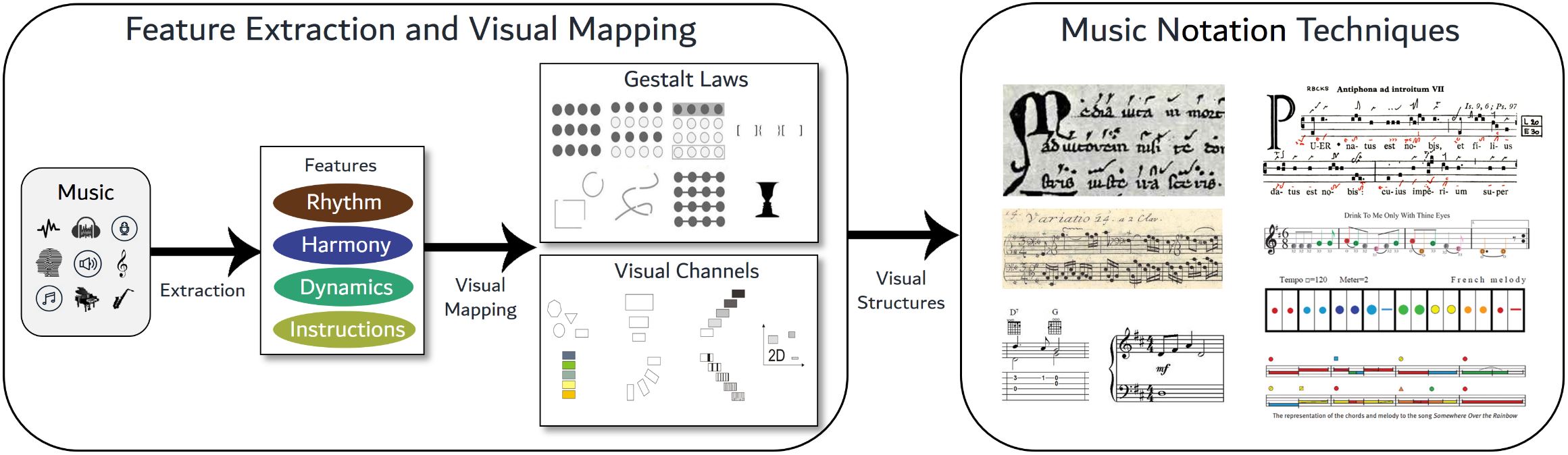}
	\caption{ 
		Segmenting music into its basic features to design and apply task-dependent visual encodings, based on visual channels and Gestalt laws. 
		Exploiting the whole range of the \textit{music notation design space} enables a variety of novel music notation techniques. 
	}
	\label{fig:teaser}
}

\usepackage{tikz} 
\usetikzlibrary{calc}
\usepackage{wasysym}
\begin{document}

	\firstsection{Introduction}
	\maketitle
	Ever since the earliest recordings of ancient cultures and societies, music has been an evident and integral part of the cultural experience and societal identity~\cite{herndon1981music}.   Musical instruments and tools have been discovered in archaeological sites, while paintings, scriptures, and folktale records form an indisputable trail that highlights the development of music through human history~\cite{sachs2012history}. However, as evident as the human desire to express feelings and emotions through music is part of our history~\cite{goldman1992value}, the preservation of the melodies of our past has been a continually changing, unguided process, influenced by many factors~\cite{lemberg2016asurvey}.

	To preserve music as part of their culture and to form a distinct identity, human societies strove to find methods that would allow them to pass such knowledge on, to later generations. Music was initially only transmitted orally~\cite{treitler1981oral}. However, due to the fading collective memories over centuries, there always existed a desire for a more accurate method of capturing and preserving musical creations~\cite{historyofmusicnotation}. Consequently, the development of the first musical notations emerged and started with a simple approach; curved lines, also called neumes~\cite{van1951musical}, were displayed over text to indicate the relative pitch for singers. Over time, the methods for music notation were altered and refined, depending on the purpose and objective of the music notation~\cite{strayer2013neumes}. This serendipitous development in different cultures and times in history led to a diversity of notation systems for various application areas and instruments.   \\
	\begin{tikzpicture}[overlay, remember picture]
	\node[anchor=north west, %anchor is upper left corner of the graphic
	xshift=2.8cm, %shifting around
	yshift=-1.4cm] 
	at (current page.north west) %left upper corner of the page
	{\includegraphics[scale=0.12]{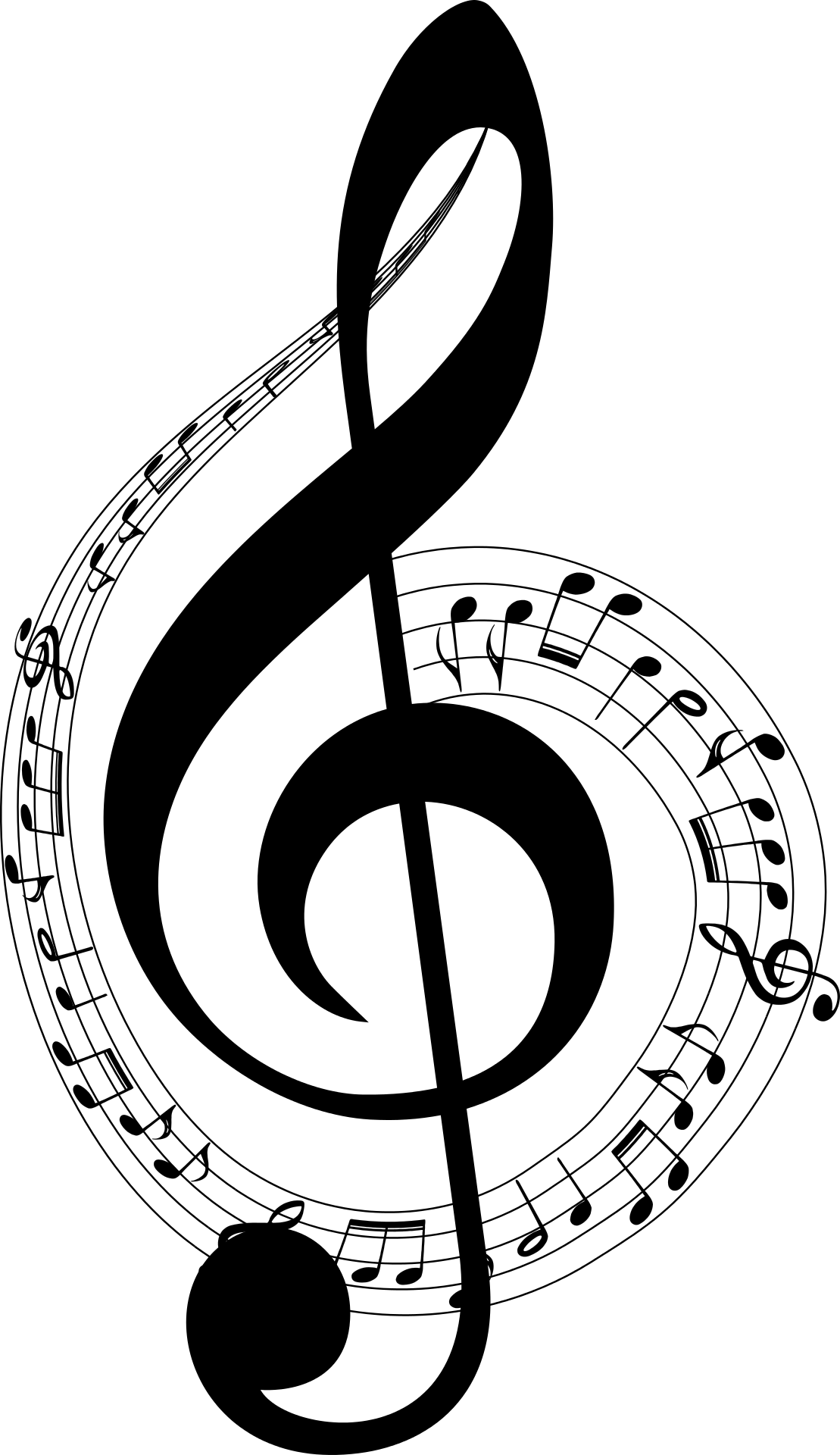}}; 
	\end{tikzpicture}
	\begin{tikzpicture}[overlay, remember picture]
	\node[anchor=north west, %anchor is upper left corner of the graphic
	xshift=17.4cm, %shifting around
	yshift=-1.3cm] 
	at (current page.north west) %left upper corner of the page
	{\includegraphics[scale=0.045]{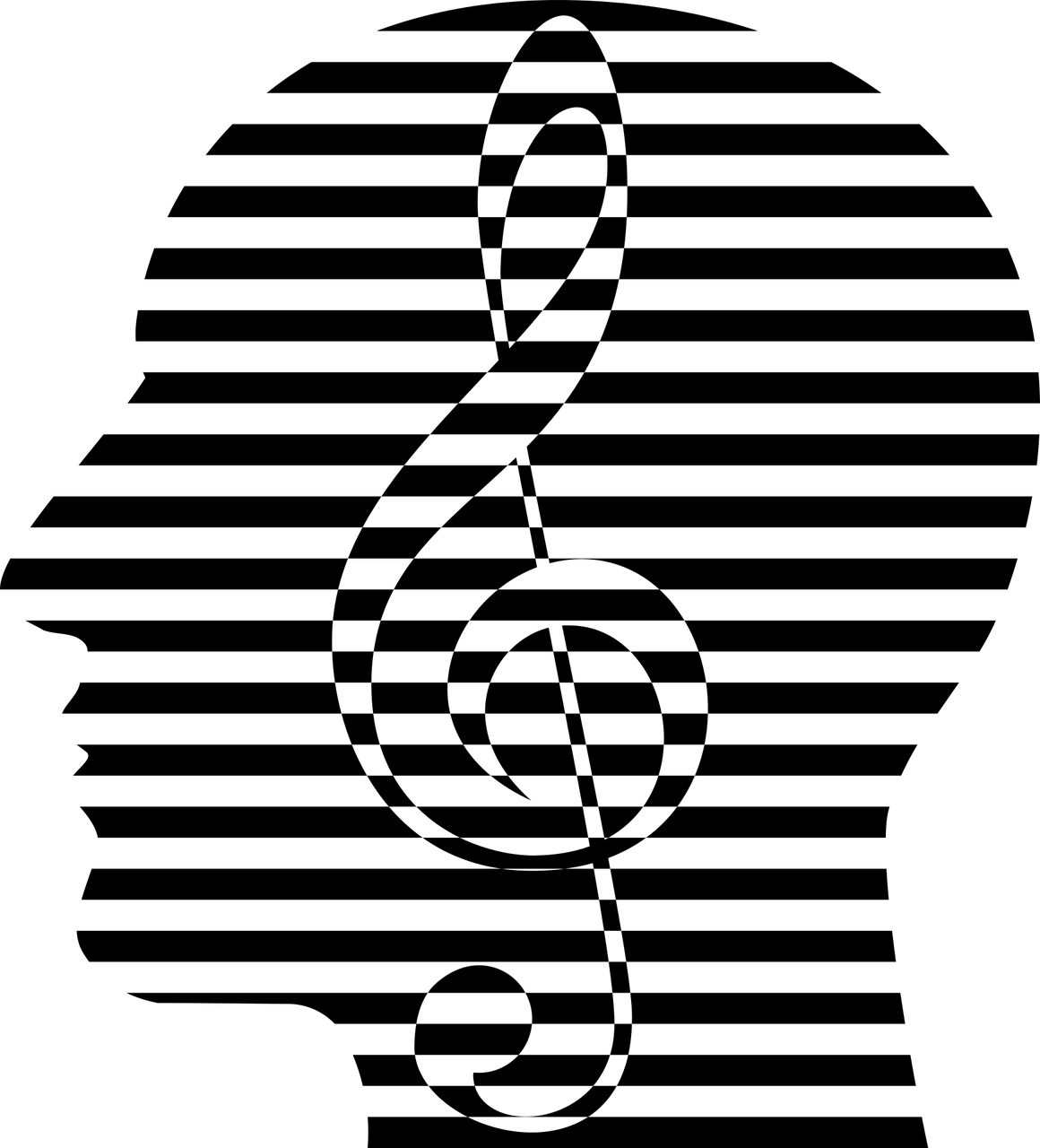}}; 
	\end{tikzpicture}
	Musical notation standardizations emerged due to the predominance of some representations, which finally led to the western, \textit{traditional} music notation that has been established as the norm we are used to all over the world. This well-known \textbf{Common Music Notation} (CMN)~\cite{Schottstaedt1997CMN} allows experienced musicians to understand how a composition should sound, solely by reading musical expressions~\cite{hultberg2002approaches}. Such expressions are commonly represented as notes on paper that are structured in a predefined notation system. Thus, one central purpose of music notation is to assist musicians with the accurate communication, interpretation, and reproducibility of compositions~\cite{chanan1995repeated}. CMN enables the capturing of musical aspects for creators and composers likewise to conserve their ideas, unambiguously. Hence, composers use music notation to communicate with performers, providing instructions on how a composition should be musically interpreted and reproduced~\cite{kendall1990communication}. However, the creation and standardization of a seemingly universal notation is challenging~\cite{dannenberg1993music}. For example, depending on the instrument, musicians use different notations~\cite{gaare1997alternatives}. Due to the varying pitch ranges between instruments, different types of clefs exist to facilitate the discrimination of music notations between instruments.  Likewise, singers require other musical representations compared to instrumentalists to understand their parts. \\

	Due to the gradual development of notation over multiple centuries, the CMN is a result of a long-term, historical development process. For instance, one bottleneck for a wide distribution was the lack of availability of printing color at the beginning of mass printing for the reproduction of music scores on paper. Similarly, the technological progress and digitization had not yet reached the point of using interactivity or motion. From the information visualization research perspective,  these historically grounded limits on the design space of music notation may have led to an established representation which does not exploit the full potential of all available techniques and encodings. For example, according to Flach et al. dyslexic individuals benefit from using an alternative notation exploiting other visual cues, enabling them to better distinguish the pitch of notes which is more difficult using the CMN~\cite{flach2016effects}. Current research in the domain of information visualization provides means for a more structured approach to the design of music notation.
	
	Hence, we postulate that {bridging the domains of information visualization and music opens up space for a structured analysis of music notation and a thorough exploration of its visual design space}. Such an endeavor opens up paths to novel and original research, tailoring music notation to different users and their tasks.

	Considering different application fields and existing music notation techniques, this paper contributes a structured analysis of the visual design space of music notation. We define the span of the design space to integrate and analyze the state-of-the-art visualization techniques of various music notations. In doing so, we obtain means to identify benefits and drawbacks of different visual encodings. In addition, our approach enables a systematic and comparative analysis of different concepts, concerning their effectiveness by applying the Gestalt laws. Finally, these considerations reveal potential research gaps and opportunities in the domain of music notation visualization.\\
    
    \vspace*{-3mm}
	
	\section{Background}
	\label{background}
	Depicting music in a visual form is a complex task containing several representation issues~\cite{dannenberg1993music}. The existence of a vast variety of different music notations shows that it is overly complicated to encode all musical features into a single and consistent system because of the limited number of visual encoding channels. Designing a notation system requires one to make fundamental design decisions that depend on the target audience and the application area. Making such design decisions limits the range of possible applications which is acceptable if some use cases are well-supported even though other tasks are not. 
	One reason why specialized notation is restricted is loss of information. 
	%Loss of information is one reason why usage of specialized notation is restricted.
	For instance, tailoring a notation system to a specific instrument makes reading more difficult for musicians who are not acquainted with the instrument. A further reason is that music is subject to a continuing progress due to the development of new instruments, genres, and even rearrangements of existing works, making it impossible to take features into account that will be introduced at a later point in time. The range of application of music notation can be divided into categories (examples are given by reference), which contain but are not limited to (live) performance~\cite{freeman2008extreme,timmers2002music}, analysis~\cite{mardirossian2007visualizing}, art~\cite{poast2000color, karttunen2016three}, education~\cite{wong2015color}, instrument support~\cite{rogers2014piano}, composition~\cite{bigo2012papertonnetz}, and entertainment~\cite{fonteles2013creating}. 
	
	Dannenberg states that the diversity between the single categories requires one to view music and its structure based on different levels of representation, since each notation design may contain information that is not available in other systems~\cite{dannenberg1993music}. 
	%He also argues that music notation systems should be extensible, since music underlies a constantly changing process prohibiting a fixed representation. 
	For instance, to analyze music, often abstract transformations are used to provide means of comparison that enable analysts to detect patterns and differences of musical pieces instead of focusing on single details~\cite{sapp2001harmonic}. Among others, music analysts are interested in understanding the harmonic progressions and relationship of a musical piece. Malandrino et al. propose a visualization approach to emphasize the harmonic structure of a composition by employing color to indicate tonal progressions~\cite{malandrino2015color}. Many approaches that propose an instrument-oriented design can also be assigned to the category `Education' since visual metaphors that are based on the visual appearance of the respective instrument provide high potential for learning a new instrument. For example, Dasc\u{a}lue et al. utilize the keyboard design to create an instrument teaching platform for adult learners~\cite{dascualu2014learning}. 
	
	To summarize the broad range of music notation systems into a single framework, we first span a \textit{Music Notation Design Space}. 
	
	\section{Music Notation Design Space}
	\label{musicnotationdesignspace}
	Visually representing music requires a suitable mapping from heterogeneous input formats to visual channels. If music exists in audio format, extracting structural features is necessary before applying a visual encoding. Since the visualization or notational representation of music is ambiguous, even existing visual representations can be examined and altered to be restructured to support different situations. Music is an intricate form of art containing both structural, as harmony or rhythm, and non-structural properties like emotion and imagination. The former is mathematically formalizable. Nevertheless, there are multiple attributes of music that hardly can be communicated using formal notation. For instance, performing music often involves a specific level of expressiveness when interpreted that is difficult to formalize or to visually encode. Consequently, dealing with music as notation entails several representational issues. This problem of representation is responsible for the existing variety of different musical representations. %Different music notation definitions address different user groups and tasks that cannot ensure lossless encoding of the original music data. 
	To analyze existing music notation specifications, we necessitate extracting musical features that are commonly used to visually represent music. Due to the hierarchical structure of music, we introduce a determined level of abstraction in subdividing music into four meta-features which comprise concrete musical dimensions each of which can be subject to visual representation in musical notation systems.

	\begin{figure*}[htp]
		\centering
		\includegraphics[width=\linewidth]{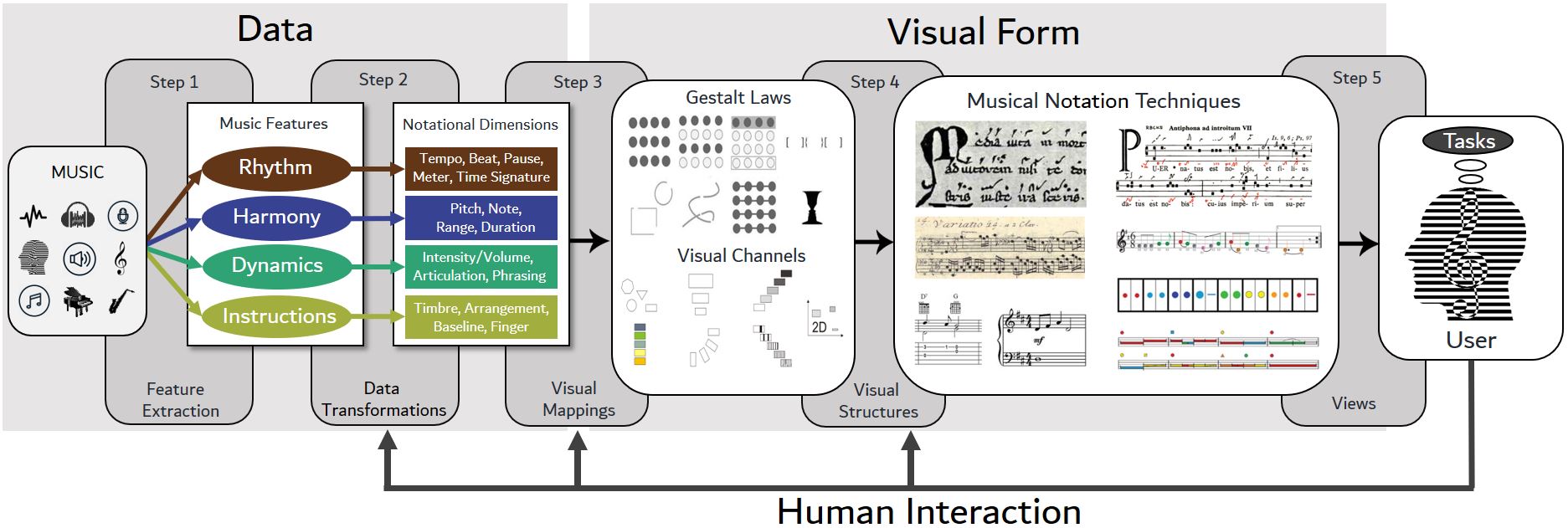}
		\caption{Our proposed \textit{Music Notation Visualization Pipeline} is based on Card et al.'s Information Visualization Reference Model \cite{card1999readings}. Transforming musical data into visual music notation as a multi-step process including extraction of musical features (Step 1), data transformation (Step 2), visual mapping (Step 3) and encoding into visual structures~(Step 4). Moreover, allowing users to interact with the system and changing the notation enables customization on views~(Step 5) and enables notation improvement to fit the users' tasks. }
		\label{fig:pipeline}
		\vspace{-10pt}
	\end{figure*}  
	
	\subsection{Musical meta-features and notational dimensions}
	\label{metafeatures}
	We suggest a set of meta-features that can be extracted from music each of which characterizes a unique musical attribute. Every meta-feature contains multiple dimensions that describe its specific musical part in detail. Subsequently, we consider music to be generally a composition of \textit{\textbf{rhythm}}, \textit{\textbf{harmony}}, \textit{\textbf{dynamics}}, and \textit{\textbf{instructions}} (see Step 1 in \autoref{fig:pipeline}). \\
	
	\noindent  \textit{\textbf{Rhythm}} describes the {tempo} (beats per minute) or speed including {meter} and {time signatures}. {Pauses} also belong to \textit{rhythm}, since they influence the rhythmic movement or flow in music. \textit{Rhythm} is the cause why music can be categorized and analyzed as time-series data and is a fundamental aspect in the design of music notation.  
	
	\noindent \textbf{\textit{Harmony}} depends on how multiple {tones} simultaneously compose specific {pitch} levels. This co-occurrence of {notes} determines the harmonic progression of a musical piece. In music notation, one must differentiate between accidental and normal notes. Moreover, the range of a given part is defined by the octaves or height of note. We assign {duration} to be part of every note since varying tone lengths directly affect the harmonic behavior. Often, dynamics is used to describe the loudness or the development of {intensity} including the transitions between different volume levels as well as accents and abrupt changes of the musical progression. 
	
	\noindent  \textbf{\textit{Dynamics}} comprises {volume (or intensity)}, {articulation} and {phrasing} since phrasing and volume are frequently combined to partition music into connected segments. Besides, {articulation} also shapes musical dynamics in combination with intensity and phrasing.  
	
	\noindent  \textbf{\textit{Instructions}} encapsulate contextual notational information such as timbre, arrangement, baseline and which finger should be used to play a specific note. `Arrangement' includes the structure of a musical piece comprising repetitions and instrumentation. Instrumentation influences the timbre and is regarded as a separate dimension which is often implicitly represented. Depending on the instrument, the CMN uses different baselines encoded by respective clefs at the beginning of the notation indicating the pitch of the displayed notes. 
	
	\subsection{Users and Tasks}Different types of users need suitable music representations: music composers must be able to expressively communicate their thoughts by symbols, signs and other instructions. They achieve this by mapping musical thoughts to be structured and represented in such a way that the composed music can be reconstructed as imagined by the author. In contrast, performers should be provided with a musical notation that is easy to understand and suitable for the instrument they are playing. The notation of a musical piece can vary between different instruments to meet task-oriented requirements. Imagine a musical director, who must be aware of all instrument parts of a whole orchestra. Providing an overview of all involved instruments in this case is beneficial. On the contrary, single instrumentalists do not profit from such an orchestra notation, but could be even distracted while playing their own part. This example emphasizes the role and importance of different music notations for diverse tasks.
	Learning an instrument is another field of application: understanding the CMN can be a challenging task. Tailoring music notation to the user while using intuitive representations such as visual instrument metaphors supports novices during the learning process.
	
	Rather abstract representations are useful for music analysts who are more interested in patterns, progressions, and general differences between musical pieces than in specific details.

	\subsection{Visual Mapping of Musical Dimensions}
	Simultaneously representing multiple features by visual cues is substantial when it comes to providing musical notation information. 
	
	During the design process of music notation, considering the \textit{Gestalt Laws} is helpful to estimate the effectiveness of a visual cue. Since visual variables are the foundation for information visualization, which musical notation is a subset of, we take Bertin's seven visual variables into account: \textit{position, size, shape, value, color, orientation}, and \textit{texture}~\cite{bertin1983semiology}. Munzner extends this list by \textit{motion, curvature, volume,} and \textit{spatial region}. Munzener divides the visual channels into \textit{magnitude channels}, which comprise ordered attributes, and \textit{identity channels} that should be used to encode rather categorical data attributes which does not have an implicit and natural ordering~\cite{munzner2014visualization}.

	It is useful to order the magnitude channels by effectiveness to reasonably apply visual channels in designing visualizations~\cite{carpendale2003considering}. Based on Munzner's discussion, we ordered the different visual magnitude channels from best to least starting with `\textit{position on common scale}' (top) to `\textit{Volume (3D size)}' (bottom) in \autoref{mappingtable}. This sorting facilitates the comparison of different visual encodings by effectiveness. Moreover, readers can get a better distribution overview of how musical features are visually encoded. 
	Applying different visual mappings between concepts avoids confusion when reading a designed music notation.

	\newcommand\Tstrut{\rule{0pt}{3.5ex}}      % "top" strut
	\newcommand\Bstrut{\rule[-2.3ex]{0pt}{0pt}}% "bottom" strut
	\newcommand{\TBstrut}{\Tstrut\Bstrut} % top&bottom struts
	
	\definecolor{yellowish}{HTML}{ffe599}
	\definecolor{brownish}{HTML}{660000}
	\definecolor{blueish}{HTML}{230085}
	\definecolor{greenish}{HTML}{38761d}
	\definecolor{darkyellowish}{HTML}{948c00}
	\definecolor{grayish}{HTML}{aaaaaa}
	\newcommand*\yellowwhite{\cellcolor{yellowish}}
	\newcommand*\brnwhitw{\cellcolor{brownish!80}\textcolor{white}}
	\newcommand*\bluewhite{\cellcolor{blueish!80}\textcolor{white}}
	\newcommand*\greenwhite{\cellcolor{greenish!80}\textcolor{white}}
	\newcommand*\darkyellowwhite{\cellcolor{darkyellowish!80}\textcolor{white}}
	
	\newcommand*\gc{\cellcolor{lightgray!70}}
	
	\newcommand*\rot{\rotatebox{90}}
	\newcommand*\rotsmall{\tiny\rotatebox{90}}
	\newcommand*\OK{\ding{51}}
	
	\newcommand{\smcite}[1]{\tiny\cite{#1}}
	
	\newcommand{\noverts}[1]{\multicolumn{1}{c}{#1}}

	\newcommand{\specialcell}[2][l]{%
		\begin{tabular}[#1]{p{0.5cm}}#2\end{tabular}}
	
	\newcolumntype{A}{>{\begin{varwidth}{0.7cm}}c<{\end{varwidth}}}
	\newcolumntype{M}{>{\begin{varwidth}{1cm}}c<{\end{varwidth}}}
	\newcolumntype{X}{>{\begin{varwidth}{1.5cm}}c<{\end{varwidth}}}
	
	\begin{table*}[t]
    
		\caption{An overview of mapping musical notation features (Rhythm, Harmony, Dynamics, Instructions) to visual variables: magnitude channel, identity channel, and Gestalt laws. The two-notes symbol (\twonotes) is used to indicate how the Common Music Notation (CMN) encodes music variables.}
		\centering
		\setlength\tabcolsep{1pt}
		\resizebox{\textwidth}{!}{
			\tiny
			\begin{tabular}{@{} c  l | MMX | MXXMX | MMM | MMMA| @{}  }
				\TBstrut
				& \noverts{{ \normalsize }}
				& \multicolumn{3}{c}{\brnwhitw{\small Rhythm}} 
				& \multicolumn{5}{c}{\bluewhite{\small Harmony}}
				& \multicolumn{3}{c}{\greenwhite{\small  Dynamics}}
				& \multicolumn{4}{c}{\darkyellowwhite{\small Instructions}} \tabularnewline
				\TBstrut	        
				& \noverts{{ \small Visual Variable}}
				& \noverts{\rotsmall{\brnwhitw{ \small Tempo / Beat}}}
				& \noverts{{\brnwhitw{  \rot{\rlap{\small Meter /}} \hspace*{0.1cm} \rot{\rlap{\small Time Signature }}  }}}  
				& \noverts{{\brnwhitw{  \rot{\rlap{\small Pauses /}} \hspace*{0.1cm} \rot{\rlap{\small  Breaks}} }}}        
				&
				\noverts{{\bluewhite{\rot{\rlap{\small Tones }} \hspace*{0.1cm}\rot{\rlap{\small (Pitch / Frequency)}}}}}
				& \noverts{\rotsmall{\bluewhite{\small Note}} }
				& \noverts{{\bluewhite{  \rot{\rlap{\small Range }} \hspace*{0.1cm}\rot{\rlap{\small (Octaves)}} }}}   
				& \noverts{{\bluewhite{  \rot{\rlap{\small Accidental / }} \hspace*{0.1cm}\rot{\rlap{\small Normal}} }}}       
				& \noverts{\rotsmall{\bluewhite{\small Duration}}}
				
				& \noverts{{\greenwhite{  \rot{\rlap{\small Intensity / }}\hspace*{0.1cm} \rot{\rlap{\small Volume}} }}}       
				& \noverts{\rotsmall{\greenwhite{\small Articulation}}}
				& \noverts{\rotsmall{\greenwhite{\small Phrasing}} }
				
				& \noverts{{\darkyellowwhite{  \rot{\rlap{\small Timbre }}\hspace*{0.1cm} \rot{\rlap{\small  (Instrument)}} }}}        
				& \noverts{\rotsmall{\darkyellowwhite{\small Arrangement}} }
				& \noverts{\rotsmall{\darkyellowwhite{\small Baseline / Clef \hspace*{0.55cm}}}}
				& \noverts{\rotsmall{\darkyellowwhite{\small Finger}}} 		   \tabularnewline        
				\cmidrule[1pt]{1-17}
				
				& Text (Semantic Channel) \cite{borgo2013glyph} 
				
				& \twonotes,  \smcite{kestenbaum2006colored}       & \twonotes, \cite{kestenbaum2006colored,rogers1996effect, de2017understanding}          &  \gc 
				
				&   \gc     &   \gc	 & 	\twonotes, \cite{kestenbaum2006colored, de2017understanding} 	  &		\gc  &		  \gc
				
				& \twonotes,  \cite{de2017understanding}  & \twonotes  &  \gc
				
				& \twonotes,  \cite{de2017understanding, cruzvisual}   &  \gc & \gc  & \twonotes, \cite{de2017understanding}    
				
				\tabularnewline
				\cmidrule{1-17}
				&  Position on common scale             
				&     \cite{nakano2005voice, ciuha2010visualization,smith1997visualization}  &   	\twonotes, \cite{kuo2013proposal, sapp2001harmonic,nakano2005voice}        &  \twonotes, \cite{kestenbaum2006colored, rogers1996effect, de2017understanding} 
				&  \cite{smith1997visualization, de2017understanding, ciuha2010visualization}       &   \cite{bergstrom2007isochords}	 & 	\twonotes, \smcite{wong2015color,kestenbaum2006colored,rogers2014piano, de2017understanding}	  &	  \gc	  &		\gc  
				&  \cite{cruzvisual}  & \gc  & \twonotes, \cite{de2017understanding}  
				&  \cite{cruzvisual, ciuha2010visualization, smith1997visualization,de2017understanding}   &  \gc  &  \gc  &  \gc  \tabularnewline
				&  Position on unaligned scale   
				&   \cite{bergstrom2007isochords}   &    \gc       &   \gc
				&    \gc    &  \twonotes, \cite{kestenbaum2006colored, de2017understanding}  	 & 		\gc  &		\gc  &		 \gc 
				&  \gc  &  \gc &  \gc
				& \gc   & \gc  &  \twonotes, \cite{kestenbaum2006colored, de2017understanding}  & \gc  \tabularnewline
				\cdashline {3-17}[1pt/1pt]
				&   Length (1D)   
				&     \gc   &         \cite{wong2015color}  &   \gc
				&   \gc     &   \gc	 & 	\cite{kuo2013proposal}	  &		\gc  &		\cite{wong2015color}  
				& \gc   & \gc  &  \gc
				&  \gc  & \gc  & \gc  & \gc \tabularnewline
				&   Tilt / Angle   
				&     \gc    &    \gc        &    \gc
				&    \gc     &    \gc	 & 	 \gc	  &	 \gc	  &		  \gc 
				&   \gc  &   \gc &   \gc
				&  \gc   &  \gc  &   \gc & \gc \tabularnewline
				\cdashline {3-17}[1pt/1pt]
				&   Area (2D size)   
				&     \gc     &       \gc      &   \gc  
				&    \gc      &   	 \gc  & 	\cite{bergstrom2007isochords}	  &		 \gc   & \cite{cruzvisual} 		  
				&  \cite{kuo2013proposal,cruzvisual}    &   \gc  &   \gc 
				&   \gc   &   \gc  &   \gc  &   \gc  \tabularnewline
				&  Depth (3D position)   
				&     \gc     &      \gc       &  \gc   
				&    \gc      &   \gc  	 & 	 \gc 	  &	 \gc   &	 \gc 	  
				&   \gc   &  \gc   &   \gc 
				&    \gc  &    \gc &  \gc   &  \gc  \tabularnewline
				\cdashline {3-17}[1pt/1pt]
				& Color Luminance   
				&      \gc    &    \cite{smith1997visualization}         &    \gc 
				&     \gc     &    \gc 	 & 	 \gc 	  &		 \gc   &		   \cite{cruzvisual}
				&  \cite{ciuha2010visualization}   &   \gc  &   \gc 
				&   \gc   &  \gc   &   \gc  & \gc  \tabularnewline
				& Color Saturation   
				&      \gc    &        \gc     &    \gc 
				&     \gc     &  \cite{ciuha2010visualization}   	 & 	 \gc 	  &		 \gc   &		 \gc   
				&   \gc   &   \gc  &   \gc 
				&   \gc   &  \gc   &   \gc  &  \gc  \tabularnewline
				\cdashline {3-17}[1pt/1pt]
				& Texture   
				&     \gc     &       \gc      &    \gc 
				&     \gc     &   \cite{wong2015color} 	 & 		 \gc   &	\cite{wong2015color}	  &		  \gc  
				&   \gc   &  \gc   &   \gc 
				&   \gc   & \cite{kuo2013proposal}    &  \gc   &  \gc                        
				\tabularnewline
				& Curvature   
				&     \gc     &       \gc      &    \gc 
				&     \gc     &  \gc   	 & 	 \gc 	  &		 \gc   &		 \gc   
				&     \gc     &       \gc      &    \gc 
				& \gc   & \gc  & \gc  & \gc \tabularnewline
				\cdashline {3-17}[1pt/1pt]
				\rot{\rlap{\small  Magnitude Channels}}
				& Volume (3D size)   
				&     \gc     &       \gc      &    \gc 
				&     \gc     &  \gc   	 & 	 \gc 	  &		 \gc   &		 \gc    
				& \cite{smith1997visualization}   &   \gc&  \gc
				&  \gc  &  \gc &  \gc & \gc  \tabularnewline
				\cmidrule{1-17}
				
				& Spatial Region  
				
				&     \gc     &       \gc      &    \gc 
				&     \gc     &  \gc   	 & 	 \twonotes 	  &		 \gc   &		 \gc   
				& \gc   & \gc  &  \gc
				&  \gc  &  \gc & \gc  & \gc \tabularnewline
				
				& Color Hue   
				
				&     \gc     &       \gc      &    \gc 
				&    \cite{sapp2001harmonic,ciuha2010visualization, toiviainen2005visualization}   & \cite{kuo2013proposal, wong2015color,kestenbaum2006colored, cruzvisual,ciuha2010visualization}    	 & 	\gc	  &	 \cite{kestenbaum2006colored}	  &		  \cite{toiviainen2005visualization}
				&  \gc  &  \gc &  \gc
				&  \cite{de2017understanding,smith1997visualization, nakano2005voice}   &  \gc & \gc  & \cite{rogers2014piano} \tabularnewline
				\cdashline {3-17}[1pt/1pt]
				
				& Motion   
				
				&    \cite{rogers2014piano,nakano2005voice,toiviainen2005visualization}     &       \cite{cruzvisual}      &    \cite{rogers2014piano}
				&     \gc     &  \gc   	 & 	 \gc 	  &		 \gc   &		 \cite{rogers2014piano,cruzvisual}   
				& \gc   & \gc  &  \gc
				&  \gc  &  \gc & \gc  & \gc \tabularnewline
				\rot{\rlap{\small Identity}}
				\rot{\rlap{\small \hspace*{-0.13cm} Channels}}
				& Shape  
				
				&    \cite{pitchbracketnotation}  &      \gc     &  \twonotes,  \cite{kestenbaum2006colored,rogers1996effect, kuo2013proposal, de2017understanding} 
				
				&    \cite{pitchbracketnotation}   &  \cite{wong2015color} 	 & 		\cite{cruzvisual}  &		\twonotes, \smcite{kuo2013proposal,kestenbaum2006colored, de2017understanding}   &	\twonotes, \cite{kuo2013proposal,kestenbaum2006colored, rogers1996effect, de2017understanding,smith1997visualization} 	  
				
				&  \gc  & \twonotes, \cite{rogers2014piano, de2017understanding}  &  \gc
				
				&  \cite{cruzvisual}  & \twonotes, \cite{kuo2013proposal, de2017understanding}   &  \twonotes, \cite{kestenbaum2006colored, de2017understanding}   & \gc \tabularnewline
				
				\cmidrule{1-17}
				
				& Proximity   
				
				&    \twonotes    &   \gc   &   \gc
				&    \twonotes, \cite{rogers2014piano}    &  \cite{cruzvisual} & 	\cite{kuo2013proposal} 	  &	\twonotes	  &		 \gc 
				&  \gc  & \twonotes  &  \gc
				&  \gc  & \gc  & \gc  & \twonotes \tabularnewline
				
				& Similarity   
				
				&  	\cite{kuo2013proposal,nakano2005voice}    &    \cite{wong2015color}     &   \twonotes
				
				&  \twonotes, \cite{cruzvisual,ciuha2010visualization}    &   \cite{kestenbaum2006colored, ciuha2010visualization, cruzvisual}	 & 	\cite{kuo2013proposal, cruzvisual,wong2015color} 	  &	\cite{kuo2013proposal,wong2015color}	  &		 \twonotes
				
				&  \gc  & \cite{rogers2014piano}   &  \gc
				
				&  \cite{de2017understanding,nakano2005voice}  & \gc  & \gc  & \gc \tabularnewline
				\cdashline {3-17}[1pt/1pt]
				
				& Enclosure   
				
				&   	\twonotes     &   \twonotes, \cite{nakano2005voice, kuo2013proposal}       & \gc  
				&     \cite{de2017understanding}   &   \gc	 &  \twonotes	  &		\gc  &		\gc  
				&  \gc  &  \gc &  \gc
				&  \gc  & \gc  &  \gc & \gc \tabularnewline
				
				& Closure   
				
				&     \gc   &   \gc       & \gc  
				&     \gc   &   \gc	 & 	\gc	  &		\gc  &		\cite{kuo2013proposal}  
				&  \gc  &  \gc &  \gc
				&  \gc  & \gc  &  \gc & \gc \tabularnewline
				\cdashline {3-17}[1pt/1pt]
				
				& Continuity   
				
				&     \cite{nakano2005voice}   &   \gc       & \gc  
				&     \twonotes   &   \gc	 & 	\gc	  &		\gc  &		\twonotes  
				&  \gc  & \cite{rogers2014piano} &  \gc
				&  \cite{de2017understanding}  & \gc  &  \gc & \gc \tabularnewline
				\rot{\rlap{\small \hspace*{-0.13cm} Gestalt Laws}}
				
				& Connection  
				
				&     \gc   &  \cite{kuo2013proposal}       & \gc  
				
				&  \cite{bergstrom2007isochords}      &   	\twonotes  & 	\cite{de2017understanding} 	  &		\gc   &	 \twonotes, \cite{wong2015color}  
				
				&  \gc  &  \cite{rogers2014piano} &  \twonotes
				
				&  \gc  & \twonotes  &  \gc & \gc \tabularnewline
				\cmidrule{1-17}
			\end{tabular}
		}
		\label{mappingtable}
		\vspace{-7pt} 
	\end{table*}

	\subsection{Music Notation Visualization Pipeline}
	We designed a pipeline to model the required transformation steps for the processing from musical data to a visual representation that is based on the Information Visualization Reference Model from Card et al. (see \autoref{fig:pipeline})~\cite{card1999readings}. In the first step, extracting musical features is required to transform the music information into a finite set of dimensions that can be potentially converted into visual form. Of course, depending on the original data format, music can contain elements such as emotions, nuances, or performance interpretation that cannot be easily visualized and conveyed to the user. During the extraction process, the loss of information should be minimized to preserve relevant information. 
	The visual mapping and the creation of visual structures is a task- and user-oriented process to meet the users' needs.
	Allowing the user to alter transformations and to influence the represented music notation provides flexibility. In some cases, users may benefit from restricted modification opportunities to maintain the quality of music representations. For example, fundamental attributes such as beat, notes, pitch, and duration should always be present. 
	Since the traditional visualization pipeline model is designed on a rather abstract level that requires the used data to be homogeneous in order to process transformation and create different views, we state that music features must be extracted before applying well-known visualization techniques. Due to the abstract level it may be require additional preprocessing steps to translate musical features to be able to process them using our proposed pipeline.

	\section{Traditional and Alternative Music Notations}
	%In the Western world, the traditional notation system has been developed over hundreds of years and is commonly used to convey musical information by musical scores. 
	Besides Common Music Notation, many different approaches have been proposed in different application domains such as research or teaching to reduce drawbacks of the CMN. These rather experimental notation concepts often use different visual channels to exploit the potential of unused visual variables. In \autoref{mappingtable}, we provide an overview of different musical notation techniques based on categories of visual channels and Gestalt laws introduced in  \autoref{musicnotationdesignspace}. Our objective is to indicate possible visual mappings that have not been used in a music notation system before, highlighted by empty cells having a grayed background. We point out that the given overview does not claim to be exhaustive since there are many ideas and approaches outside the world of academia~ (e.g., Pitch Bracket Notation~\cite{pitchbracketnotation}). In the table, the CMN is listed using a two notes symbol (\twonotes) to emphasize differences between the listed notation approaches. Moreover, we selected the included references to be reproducible by a performer or computer instead of adding all methods that visualize any musical feature for analysis or entertainment. Frequently, music notation analysis visualizations apply abstract methods that do not take fundamental notation features into account\cite{li2016music, wattenberg2002arc}. 
	
	In \autoref{mappingtable} we classify fifteen existing music notation techniques. We emphasize that this is not representative for all notation techniques that exist in literature. It is structured to facilitate the understanding of which reference of a music representation method is using a visual variable (rows) grouped by Munzners visual channel categorization \cite{munzner2014visualization} and Gestalt law to encode a musical feature (columns). The columns are grouped by the characteristic musical meta-features introduced in \autoref{metafeatures}. Some music notation concepts use different encodings for notes (a tone within an octave: A-G), ranges, and pitch. We consider pitch to be composed by note and octave. Mostly, a performer requires the exact position to precisely play a musical piece. Therefore, we decided to list them separately to highlight explicit representations of this musical dimension. Music notation often includes textual descriptions to provide information about instrumentation, volume, tempo, and meter signature. Borgo et al. subdivide all visual channels into four categories, one of which are the \textit{Semantic Channels} containing text, number, symbols, signs, icons, and others \cite{borgo2013glyph}. We consider this category to be appropriate to describe the contextual music notation information (first row in the overview table).
	Seldomly, complete chords are directly mapped on visual variables instead of the respective single notes. For instance, Malandrino et al. enrich the CMN with a colored background to indicate the current pitch class at a particular position in the score \cite{malandrino2015color}. For the sake of simplicity and clarity, we decided to omit a separate column for the chord dimension due to the scarcity of such techniques.\\
	
	\section{Research Implications and Opportunities}

	Our research shows how notations can be compared regarding their visual features. 
	To the best of our knowledge, \autoref{mappingtable} comprises varying music notations but is not complete and may serve as a starting point for a survey. 
	The introduced visualization pipeline (Fig.~\ref{fig:pipeline}) helps to both understand existing techniques and develop new ones. Since the pipeline unifies the design process on a rather abstract and conceptual level, it may be required to consider some musical dimensions in more detail. Nevertheless, we claim that any music notation visualization can be modeled using our pipeline that can be extended to cover further musical dimensions that we did not consider, if necessary. In doing so, weaknesses of a specific representation, as well as its capability to visually map musical features for a given task, can be revealed, considering relevant notational dimensions. Vice versa, in the design process of new music notation visualizations, the pipeline in combination with the overview table supports the consideration of different notational dimensions.	The line-up of mappings between musical features and visual variables can also be used as inspiration to create new higher-level visualizations which also support content-based visual analysis of music. 
	
	Our research can be seen as a starting point for in-detail investigations of certain combinations of mappings and users or tasks. Specific pairs of visual variables and musical features could particularly fit certain user groups (e.g., learners, composer) or tasks (e.g., high-level analysis of a musical piece, performance).
	We assume that readers of the classical musical notation system are inherently biased due to early familiarization with the standard notation. Due to the familiarity, the popularity of the CMN and the historical contingency, it can be a difficult process to develop a music notation that will finally replace the CMN.

	By now, we can exploit visual channels requiring technology that was not available during the development process of the CMN such as motion or color. 
	Since CMN has some disadvantages such as differentiating tone pitches, new techniques could address these drawbacks by applying visual variables in a different way. 
	During the design process of new music visualization methods, it is necessary to take the respective application area into account, as some mappings are more intuitive in a specific situation and convey precise while others rather provide abstract information of musical features.

	We argue that our design and visualization pipeline for music notation (see \autoref{fig:pipeline}) is also applicable to develop methods for music fingerprinting to compare music regarding different epochs and styles, or to compare the visual encoding of music visualizations.
	For instance, a well-established mapping for some notational feature could be deployed in a higher level visualization, showing, for example, the trend of a musical feature throughout the entire musical piece. Visualization techniques as pixel visualizations, glyphs, or others could be applied to visually analyze larger amounts of musical data and enable comparison between musical pieces. 
	
	\section{Conclusion}
	We introduced a design and visualization pipeline for music notation based on fundamental information visualization research. This approach can be used to compare existing music notation designs and to develop new techniques that are tailored to different users and their tasks. Thus, we bridge the research field of data visualization and the field of music representation in a structured way. An exemplary classification of fifteen  notation systems indicates how musical features can be encoded. The table can be used as a starting point for a full survey of existing music notation techniques.

	\bibliographystyle{abbrv-doi-hyperref}
	\bibliography{template}
\end{document}